\documentclass[a4paper,11pt]{article}
\usepackage{pos}

\usepackage{xcolor}
\usepackage{graphicx}
\usepackage{wrapfig}
\usepackage{amsmath}
\usepackage{multirow}
\usepackage{slashed}
\usepackage{physics}
\usepackage{booktabs}
\usepackage{bbm}

\newcommand \ua {U(1)_{\mathrm A}}
\newcommand \ml {m_l}
\newcommand \ms {m_s}
\newcommand \tc {T_c}
\newcommand \lda {\lambda}
\newcommand \ro {\rho(\lda, \ml)}
\newcommand \ru[1] {\rho_U(\lda_{#1})}
\newcommand \av[1] {\left\langle{#1}\right\rangle}
\newcommand \U {\mathcal{U}}
\newcommand \nt {N_\tau}
\newcommand \ns {N_\sigma}
\newcommand \cpi {\chi_\pi}
\newcommand \cdl {\chi_\delta}
\newcommand \cdsc {\chi_\mathrm{disc}}

\newcommand \pbp {\langle \bar{\psi}\psi\rangle}
\newcommand{\su}{SU(2)_\mathrm{L}\times SU(2)_\mathrm{R}}

\title{Correlated Dirac Eigenvalues and Axial Anomaly in Chiral Symmetric QCD}

\author[1]{H.-T. Ding}
\author[1]{S.-T. Li}
\author[2]{Swagato Mukherjee}
\author[3]{A. Tomiya}
\author[1]{X.-D. Wang}
\author*[4]{Y. Zhang}

\affiliation[1]{Key Laboratory of Quark \& Lepton Physics (MOE) and Institute of Particle Physics,\\
  Central China Normal University, Wuhan 430079, China}

\affiliation[2]{Physics Department, Brookhaven National Laboratory, Upton, NY 11973, USA}

\affiliation[3]{RIKEN-BNL Research Center, Brookhaven National Laboratory, Upton, New York 11973, USA}

\affiliation[4]{RIKEN Center for Computational Science, 7-1-26	Minatojima-minami-machi, Chuo-ku, Kobe, Hyogo 650-0047, Japan}

\emailAdd{hengtong.ding@mail.ccnu.edu.cn}
\emailAdd{lishengtai@mails.ccnu.edu.cn}
\emailAdd{akio.tomiya@riken.jp}
\emailAdd{swagato@bnl.gov}
\emailAdd{xiaodanwang@mails.ccnu.edu.cn}
\emailAdd{yu.zhang.ey@riken.jp}

\abstract{ We investigate the Dirac eigenvalue spectrum ($\ro$) to study the microscopic origin of axial anomaly in high temperature phase of QCD. We propose novel relations between the derivatives ($\partial^n{\ro}/\partial\ml^n$) of the Dirac eigenvalue spectrum with 
respect to the quark mass ($\ml$) and the $(n+1)$-point correlations among
the eigenvalues ($\lda$) of the massless Dirac operator. Based on these relations, we present lattice QCD results for
$\partial^n{\ro}/\partial\ml^n$ ($n=1, 2, 3$) with $\ml$ corresponding to pion masses
$m_\pi=160-55$~MeV, and at a temperature of about 1.6 times the chiral phase
transition temperature. Calculations were carried out using (2+1)-flavors of highly
improved staggered quarks and the tree-level
Symanzik gauge action with the physical strange quark mass, three
  lattice spacings $a=0.12, 0.08, 0.06$~fm, and lattices having aspect ratios $4-9$. We find that $\rho(\lda\to0,\ml)$ develops a peaked structure. This peaked structure, which arises due to non-Poisson correlations within the infrared part of the
Dirac eigenvalue spectrum, becomes sharper as $a\to0$, and its amplitude is proportional to $\ml^2$. After continuum and chiral extrapolations, we find that the axial anomaly remains manifested in two-point correlation functions of scalar and pseudo-scalar mesons in the chiral limit. We demonstrate that the behavior of $\rho(\lda\to0,\ml)$ is responsible for it. } 

\FullConference{%
 The 38th International Symposium on Lattice Field Theory, LATTICE2021
  26th-30th July, 2021
  Zoom/Gather@Massachusetts Institute of Technology
}

\begin{document}
\maketitle

\section{Introduction}
The Lagrangian of the (2+1)-flavor Quantum Chromodynamics (QCD) has a global symmetry $\su \times \ua \times U(1)_V$ in the classic limit and the chiral limit of $\ml\to0$.  The $\su$ chiral symmetry is spontaneously broken in the vacuum and the
$\ua$ symmetry is anomalously broken on the quantum level due to the Adler-Bell-Jackiw or chiral anomaly. 
For the physical $\ml$ lattice simulations have established quite firmly that QCD transition is a rapid cross over at a pesudocritical temperature at $T\simeq156$~MeV~\cite{Aoki:2009sc, Bhattacharya:2014ara,
Bazavov:2018mes}, while in the chiral limit $\ml\to0$ chiral phase transition temperature at which the $\su$ is restored is estimated as 
$\tc=132^{+3}_{-6}$~MeV based on the O(4) scaling analyses~\cite{Ding:2019prx}.

Conversely, the fate of the $\ua$ symmetry in the high temperature phase of QCD remains unclear. Although the quantum anomaly is present at any finite temperature, at some point its effects could become negligible due to the asymptotic restoration of the $\ua$ symmetry with the temperature, thus the $\ua$ symmetry would be effectively restored.
The order of the chiral transition and the associated universality class is known to depend crucially on how axial anomaly manifests itself in the two-point correlation
functions of light scalar and pseudoscalar mesons for $T\ge\tc$. If the isotriplet
scalar $\delta$ and the isotriplet pseudoscalar $\pi$ remain non-degenerate at
$T\ge\tc$, then we expect a second order phase transition which 
belongs to the three-dimensional $O(4)$ universality class~\cite{Pisarski:1983ms}. But if the $\delta$
and $\pi$ become degenerate at $T\ge\tc$, then 
the chiral phase transition can be either first~\cite{Pisarski:1983ms} or
second order with the symmetry breaking pattern $U(2)_V \times U(2)_A \to U(2)_V$ universality class~\cite{Butti:2003nu, Grahl:2014fna}. For the physical
$\ml$, the $\delta$ and  $\pi$ remain nondegenerate around the chiral
crossover~\cite{Buchoff:2013nra, Bhattacharya:2014ara,
Bazavov:2019www}. However, what happens for $T\simeq\tc$ as $\ml\to0$ remains an open
question~\cite{Ohno:2012br,Dick:2015twa,
Tomiya:2016jwr, Aoki:2020noz} due to the lack of state-of-the-art
lattice QCD calculations with controlled continuum and chiral extrapolations.

To gain more insight about the microscopic origin of the axial anomaly we can investigate the Dirac eigenvalue spectrum $\ro$.
It has been shown that if $\ro$ is an analytic function of $\ml^2$ and $\lda$ then in the chiral limit $\ua$ breaking effects are invisible in
differences of up to 6-point correlation functions of $\pi$ and $\delta$ that can be
connected via a $\ua$ rotation~\cite{Aoki:2012yj}. However,
the dilute instanton gas approximation (DIGA)~\cite{tHooft:1976rip}  predicated that $\rho \sim \ml^2
\delta(\lda)$ can lead to nondegeneracy of the two-point $\pi$ and $\delta$
correlation functions even as $\ml\to0$~\cite{Bazavov:2012qja,Gross:1980br,Kanazawa:2014cua}. Some lattice QCD studies have
observed infrared enhancement in $\rho$~\cite{Bazavov:2012qja, Buchoff:2013nra,
Dick:2015twa}, however, whether such enhancements scale as $\ml^2$ as $\ml\to0$ have
not been demonstrated. In other lattice QCD calculations, no infrared enhancement in
$\rho$ was observed~\cite{Cossu:2013uua, Tomiya:2016jwr,
Suzuki:2020rla}, showing the importance of controlling lattice artifacts through
continuum extrapolations. On the other hand,  in Ref.~\cite{Kanazawa:2015xna} it was
argued that if $\pi$ and $\delta$ were to remain nondegenerate at $T\ge\tc$,
then chiral symmetry restoration demands non-Poisson correlations among the infrared
eigenvalues.

In this work we propose the novel relation between $\partial^n \rho/\partial m_l^n$ and correlation among the eigenvalues to investigate the microscopic origin of axial anomaly at high temperature phase. The rest of paper is organized as follows. We describe the basic idea of how to obtain the relation between $\partial^n \rho/\partial m_l^n$ and correlation among the eigenvalues in section~\ref{2}. In section~\ref{3} we show the setup of our lattice simulations. We then show our numerical results in Section~\ref{4}. Finally we present our conclusion in Section~\ref{5}. The detailed information about this work can be found in~\cite{Ding:2020xlj}.

\section{\texorpdfstring{$\partial^n \rho/ \partial m_l^n \,\& \,C_{n+1}$}{∂\^nρ/∂m\_l\^n} and \texorpdfstring{$U(1)_A$}{U(1)\_A} anomaly}
\label{2}
For (2+1)-flavor QCD, the Dirac eigenvalue spectrum is given by
\begin{equation}
  \label{eq:ro}
  \begin{split}
    \ro = \frac{T}{V Z[\U]} &
    \int \mathcal{D}[\U] e^{-S_G[\U]} \det\qty[\slashed{D}[\U]+\ms]  \qty(\det\qty[\slashed{D}[\U]+\ml])^2 \ru{} \,.
  \end{split}
\end{equation}
Here $\ru{}$ is the Dirac eigenvalue spectrum for a given gauge configuration, It is defined as $\ru{} = \sum_j \delta(\lda-\lda_j)$, $\lda_j$ are the eigenvalues of the
massless Dirac matrix $\slashed{D}[\U]$. Note that $\ru{}$ does not explicitly depend
on $\ml$ and the $\ml$ dependence is embedded in the determinant term.
Furthermore,
\begin{equation}
  \label{eq:det}
  \begin{split}
    & \det\qty[\slashed{D}[\U]+\ml] = \prod_j \qty(+\mathrm{i}\,\lda_j+\ml) \qty(-\mathrm{i}\,\lda_j+\ml) 
    = \exp \qty( \int_0^\infty \dd{\lda} \ru{} \ln\qty[\lda^2+\ml^2] ) \,.
  \end{split}
\end{equation}
Substituting \autoref{eq:det} in \autoref{eq:ro} and $Z[\U]$ it is straightforward to
obtain $\partial^n \rho/\partial m_l^n$ ~\cite{Ding:2020xlj}, \text{e.g.,}
\begin{align}
  \label{eq:dro1}
  & \frac{V}{T} \frac{\partial \rho}{\partial \ml} = \int_0^\infty {\rm d}{\lda_2}
  \frac { 4 \ml\, C_2(\lda, \lda_2; \ml)} {  \lda_2^2 + \ml^2 } \,, \\
  \label{eq:dro2}
  \begin{split}
    & \frac{V}{T} \frac{\partial^2 \rho}{\partial \ml^2} =
   \int_0^\infty \dd{\lda_2}
    \frac { 4 (\lda_2^2-\ml^2) \,C_2(\lda, \lda_2; \ml) }
    { \qty( \lda_2^2 + \ml^2 )^2}
    + \int_{0}^{\infty}  d\lambda_3   \int_{0}^{\infty}  d\lambda_2
    \frac { (4\ml)^2\, C_3(\lda, \lda_2,\lda_3; \ml) }
    { \qty( \lda_2^2 + \ml^2 ) \qty( \lda_3^2 + \ml^2 ) }
    \,,
  \end{split} \\
   \begin{split}
  \label{eq:Cn}
  & \qq{with}  C_n(\lda_1, \cdots, \lda_n; \ml) =
  \av{\prod_{i=1}^n \qty[ \ru{i} - \av{\ru{i}} ] } .
\end{split}
    \end{align}
The difference of the integrated two-point functions in the pion and delta channel is defined as 
\begin{equation}
  \label{eq:cdif}
  \cpi - \cdl = \int \dd[4]{x} \av{ \pi^i(x)\pi^i(0) - \delta^i(x)\delta^i(0) } \,.
\end{equation}
For $T \ge \tc$ owing to the degeneracy of $\pi$ and the $\sigma$ in
the chiral limit~\cite{Bazavov:2012qja}
\begin{equation}
  \label{eq:rel}
  \cpi - \cdl = \cdsc \,,
\end{equation}
where $\cdsc$ is the quark-line disconnected part of the isosinglet scalar meson
susceptibility,
\begin{equation}
  \label{eq:cdsc}
  \cdsc = \frac{T}{V} \int \dd[4]{x} \av{ \qty[ \bar\psi(x)\psi(x)
   - \av{\bar\psi(x)\psi(x)} ]^2 }  \,.
\end{equation}
The $\mathrm{\ua}$ symmetry-breaking measures $ \cpi - \cdl$ and $\cdsc$ are related to $\rho$
through~\cite{Bazavov:2012qja,Ding:2020xlj}
\begin{align}
  \label{eq:sus-rho}
   \cpi - \cdl = \int_0^\infty \dd{\lda} \frac { 8 \ml^2 \,\rho } { \qty( \lda^2 + \ml^2
  )^2 } \,,\qquad
   \cdsc = \int_0^\infty \dd{\lda} \frac { 4 \ml\,\pdv*{\rho}{\ml}} { \lda^2 +\ml^2  } \,.
\end{align}
In the Poisson limit, $C_n^{\mathrm{Po}}(\lda_1, \cdots, \lda_n) =
\delta(\lda_1-\lda_2) \cdots \delta(\lda_n -\lda_{n-1})
\av{\qty(\ru{1} -\av{\ru{1}})^n} = \delta(\lda_1-\lda_2) \cdots
\delta(\lda_n-\lda_{n-1}) \av{\ru{1}} + \order{1/N}$, where $2N\propto V/T$ is the total
number of eigenvalues. In this limit the first and second order quark mass derivatives of $\rho$ are expressed as follows
\begin{align}
  \label{eq:droPo}
  & \qty(\pdv{\rho}{\ml})^{\mathrm{Po}} = \
  \frac{4\ml\rho}{\lda^2+\ml^2} - \frac{V\rho}{TN} \av{\bar\psi\psi} \,,\\
  \label{eq:dro2Po}
  & \qty(\pdv[2]{\rho}{\ml})^{\mathrm{Po}} = \frac{4\rho}{\lda^2+\ml^2} +
  \frac{8\ml^2\rho}{\qty(\lda^2+\ml^2)^2} + \frac{2V^2\rho}{T^2N^2} \av{\bar\psi\psi}^2
 - \frac{V\rho}{TN} \qty( \frac{8\ml\av{\bar\psi\psi}}{\lda^2+\ml^2} + 2\cpi - \cdl ) \,,
\end{align}
In the chiral limit, this
leads to $\cdsc^\mathrm{Po} = 2(\cpi-\cdl)$, in clear violation of the chiral
symmetry restoration condition in \autoref{eq:rel}, unless both sides of the equation
trivially vanish.
\section{Lattice setup}
\label{3}

Lattice QCD calculations were carried out at $T \approx   205$~MeV $\approx1.6\tc$ for
$(2+1)$-flavor QCD using the highly improved staggered quarks and the tree-level
Symanzik gauge action. The $\ms$ was tuned to its physical value, and three lattice
spacings $a = (T\nt)^{-1} = 0.12, 0.08, 0.06$~fm corresponding to $\nt=8, 12, 16$, were used~\cite{Ding:2020xlj}. Calculations were done
with $\ml = \ms/20, \ms/27, \ms/40, \ms/80, \\
\ms/160$ that correspond to $m_\pi \simeq
160, 140, 110, 80, 55$~MeV, respectively. The spatial extents ($\ns$) of the lattices
were chosen to have aspect ratios in the range of $\ns/\nt=4-9$. $\rho$ and $C_n$ were computed by measuring  $\ru{}$ over the
entire range of $\lda$ using the Chebyshev filtering technique combined with the
stochastic estimate method~\cite{Ding:2020eql, Giusti:2008vb, Cossu:2016eqs,Ding:2020hxw} on about $2000$ configurations where each configuration is separated by 10 time units. Orders of the Chebyshev
polynomials were chosen to be $(1-5)\times10^5$ and 24 Gaussian stochastic sources
were used. Measurements of $\cdsc$ and $\cpi-\cdl$ were done by inverting the light
fermion matrix using $50$ Gaussian random sources on $2000-10000$
configurations. Apart from the data sets as shown in above which were reported in~\cite{Ding:2020xlj}, in this paper we also add new results based on simulations with $\ml=\ms/160$ on $\nt=12,16$ lattices. For each of these two parameter sets 4200-5200 configurations each separated by 10 time units are generated, and $\cdsc$ and $\cpi-\cdl$ are measured by inverting the fermion matrix on these configurations.

\section{Results}
\label{4}

\begin{figure*}[!htp]
  \centering
	\includegraphics[width=0.32\textwidth,height=0.16\textheight]{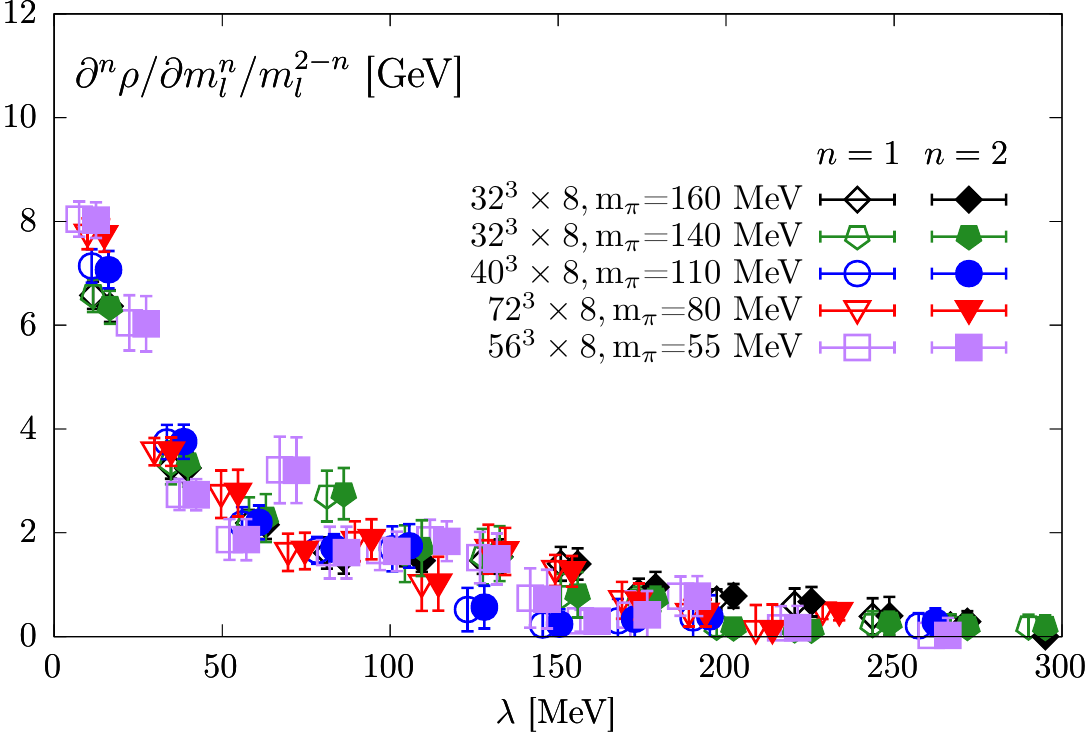}
	\includegraphics[width=0.32\textwidth,height=0.16\textheight]{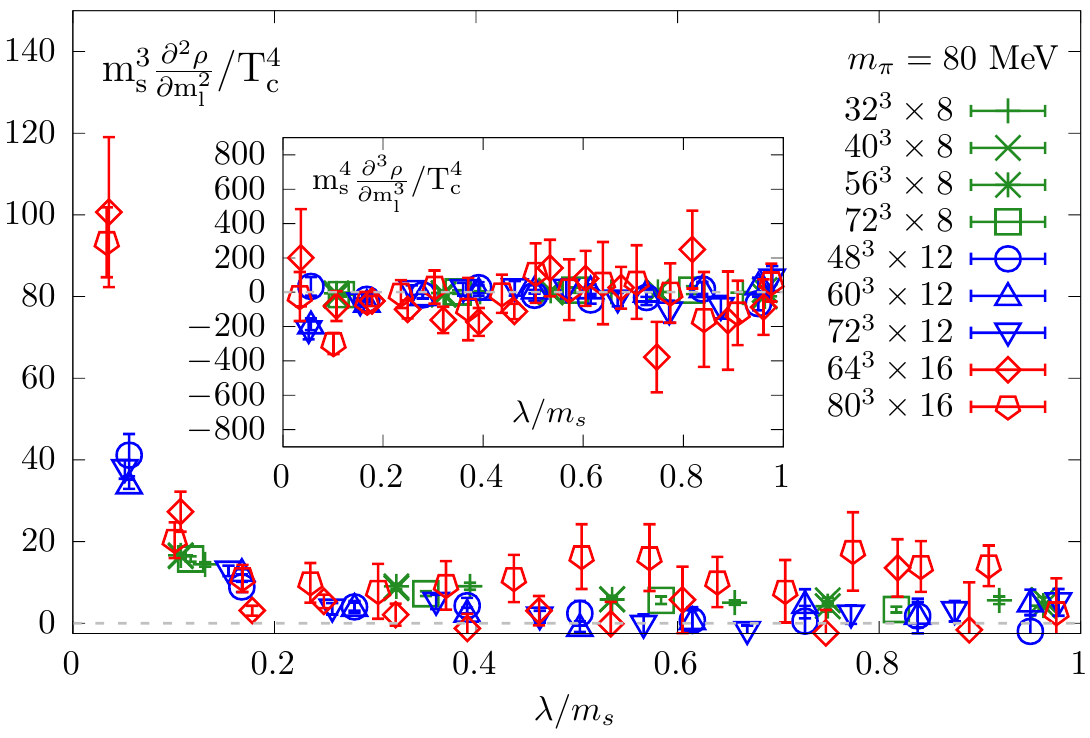}
        \includegraphics[width=0.32\textwidth,height=0.16\textheight]{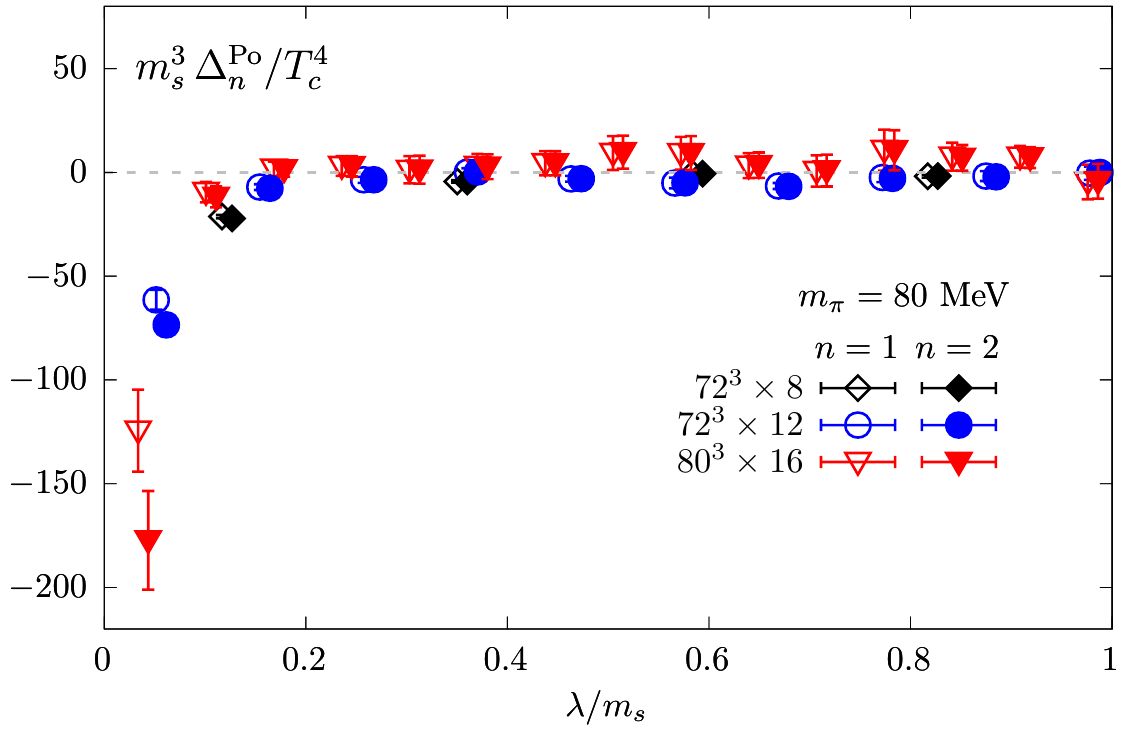}
\caption{Left: $\ml$ dependence of $m_l^{-1}\pdv*{\ro}{\ml}$ and $\pdv*[2]{\ro}{\ml}$  using $\nt=8$ lattices. Middle: $a$ and $V$ dependence of $\pdv*[2]{\ro}{\ml}$ and
	$\partial^3 \ro/\partial m_l^3$ (inset) for $m_\pi=80$~MeV. Right: The differences
	$\Delta_n^\mathrm{Po} = \ml^{n-2} \qty[ \partial^n \rho/\partial m_l^n -
	(\partial^n \rho/\partial m_l^n )^\mathrm{Po} ]$ for $m_\pi=80$~MeV and three lattice spacings.}
	\label{fig:rho}
\end{figure*}
\autoref{fig:rho} (left) shows the $\ml$ dependence of $m_l^{-1}\partial \rho/\partial m_l$ and
$\partial^2 \rho/\partial m_l^2$ at $T\approx1.6\tc$, obtained on $\nt=8$ and the
largest available $\ns$ for that $\ml$. We observe that $\ml^{-1}(\partial \rho/\partial m_l)$
and $\partial^2 \rho/\partial m_l^2$ are almost equal to each other and independent of $\ml$. Also,
$\ml^{-1}\partial \rho/\partial m_l$ and  $\partial^2 \rho/\partial m_l^2$ develops a peak at $\lda\to0$ and it drops
rapidly toward zero for $\lda/T\gtrsim1$. \autoref{fig:rho} (middle) depicts the
lattice spacing and volume dependence of $\partial^2 \rho/\partial m_l^2$ and
$\partial^3 \rho/\partial m_l^3$ for $m_\pi=80$~MeV. To compare these quantities across
different lattice spacings we multiply with the appropriate powers of $\ms$ to make
them renormalization group invariant and make them dimensionless by rescaling
with appropriate powers of $\tc=132$~MeV. We see that the  peaked structure in
$\partial^2 \rho/\partial m_l^2$ at $\lda\to0$ becomes sharper as $a\to0$, and shows little
volume dependence. Moreover, $\partial^3 \rho/\partial m_l^3$ are found to be
consistent with zero within errors. 
The findings
$\ml^{-1}\partial \rho/\partial m_l\approx\partial^2 \rho/\partial m_l^2$ and
$\partial^3 \rho/\partial m_l^3 \approx0$ show that the peaked structure
$\rho(\lda\to0,\ml\to0)\propto\ml^2$. In \autoref{fig:rho} (right) we show the difference
	$\Delta_n^\mathrm{Po} = \ml^{n-2} \qty[ \partial^n \rho/\partial m_l^n -
	(\partial^n \rho/\partial m_l^n )^\mathrm{Po} ]$ ($n=1,2$), with the Poisson approximations for
$\partial^n \rho/\partial m_l^n$ as defined in \autoref{eq:droPo} and \autoref{eq:dro2Po}. The
fact $\Delta_n^\mathrm{Po}<0$ shows that the repulsive non-Poisson correlation
within the small $\lda$ gives rise to the $\rho(\lda\to0)$ peak.

\begin{figure}[!htp]
  \centering
	\includegraphics[width=0.495\textwidth, height=0.25\textheight]{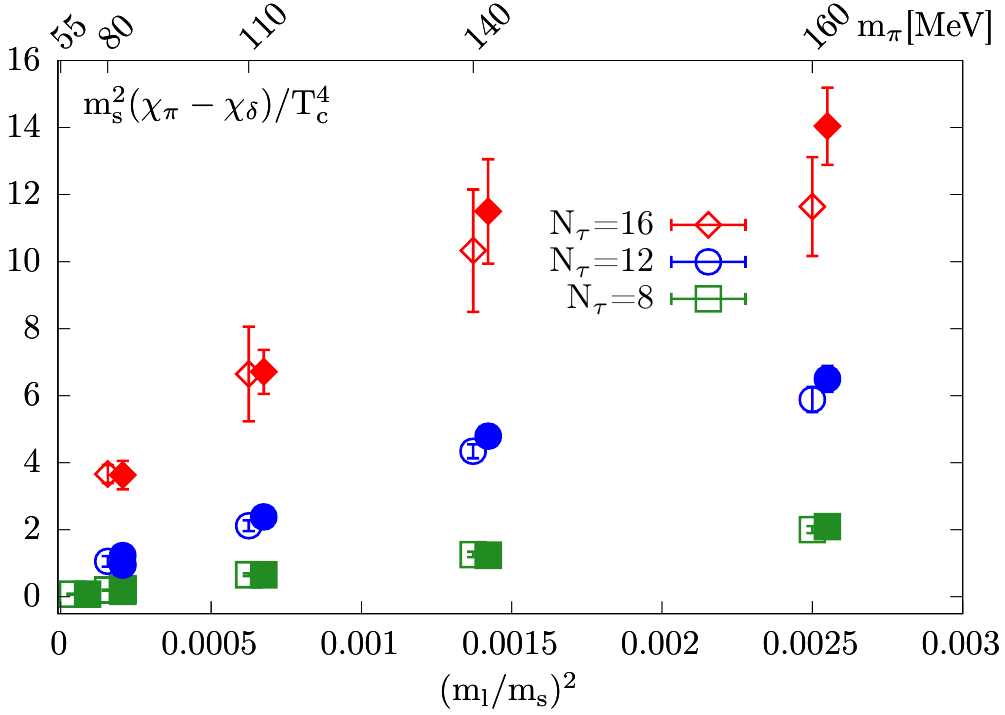}
	\includegraphics[width=0.495\textwidth, height=0.25\textheight]{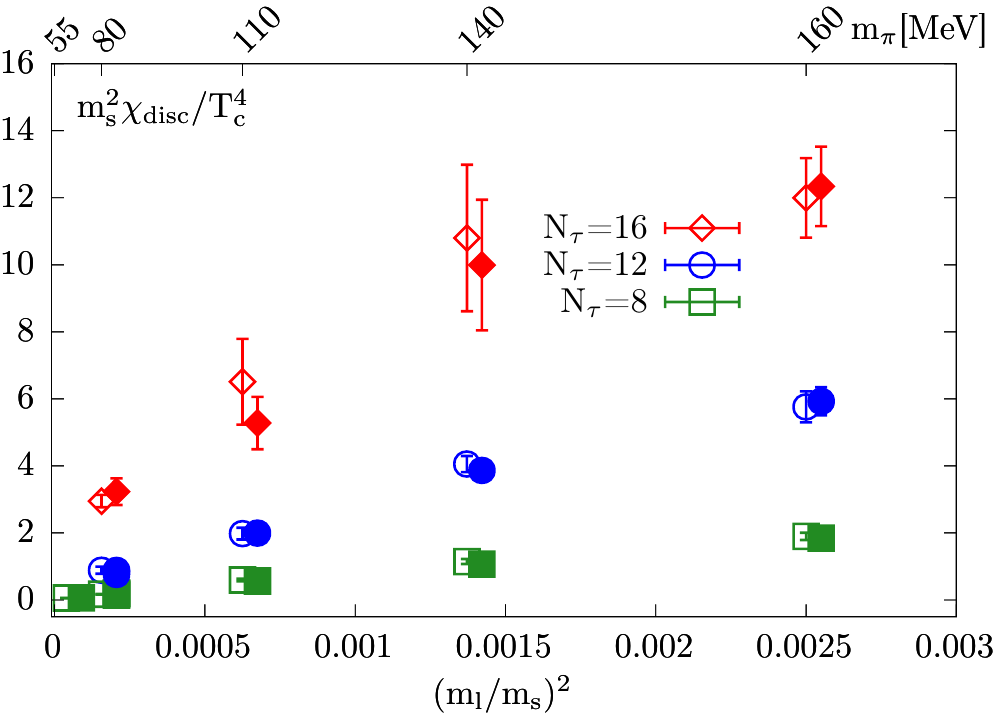}
	\caption{Comparisons of direct measurements (open symbols) of $\cpi-\cdl$ (left) and
	$\cdsc$ (right) with those reconstructed (filled symbols, slightly shifted
	horizontally for visibility) from $\rho$ and
	$\partial \rho/\partial m_l$ using \autoref{eq:sus-rho}.} 
  \label{fig:Comparison}
\end{figure}

In \autoref{fig:Comparison} we show that $\rho$ and $\partial \rho/\partial m_l$ reproduce
directly measured $\cpi-\cdl$ and $\cdsc$ using \autoref{eq:sus-rho}. 
We checked that only the infrared $\lda/T\lesssim1$ parts of $\rho$
and $\partial \rho/\partial m_l$ are needed for the reproductions of $\cpi-\cdl$ and $\cdsc$. Additionally, we checked that once the bin-size of $\lda$ in the numerical integration of left equation of \autoref{eq:sus-rho} is chosen to
reproduce directly measured $\cpi-\cdl$, the same bin size automatically reproduces
$\cdsc$ and $\av{\bar\psi\psi}$ without any further tuning. 
\begin{figure*}[!htp]
\centering
	\includegraphics[width=0.51\textwidth, height=0.228\textheight]{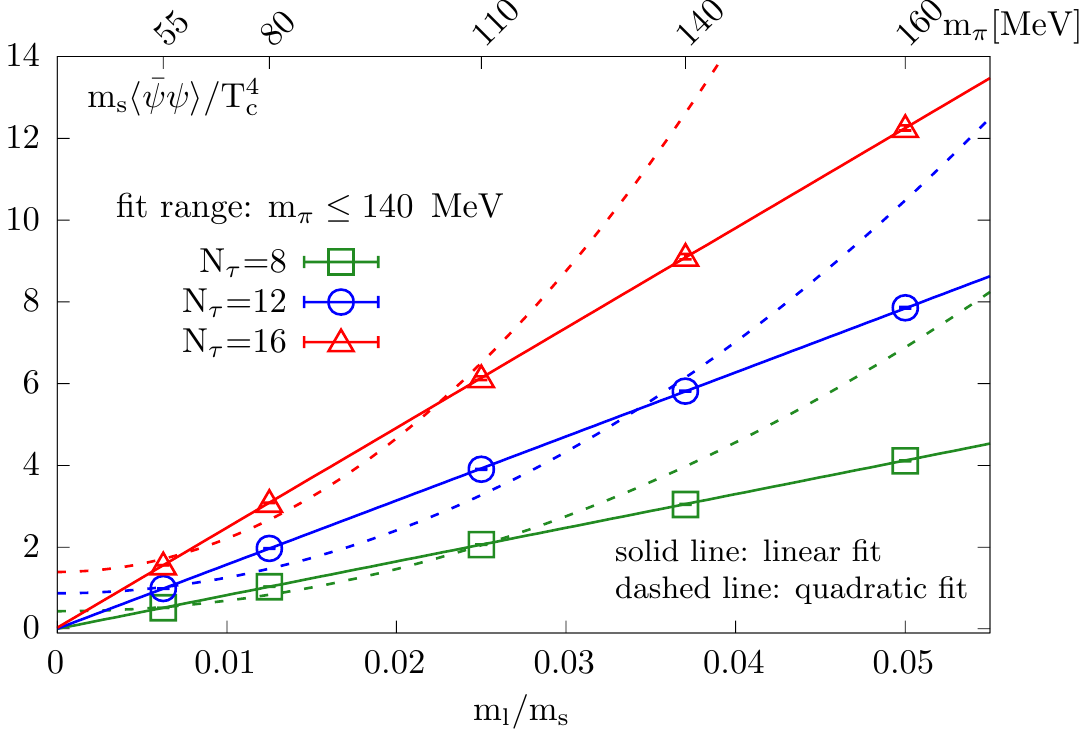}~
		\includegraphics[width=0.48\textwidth, height=0.235\textheight]{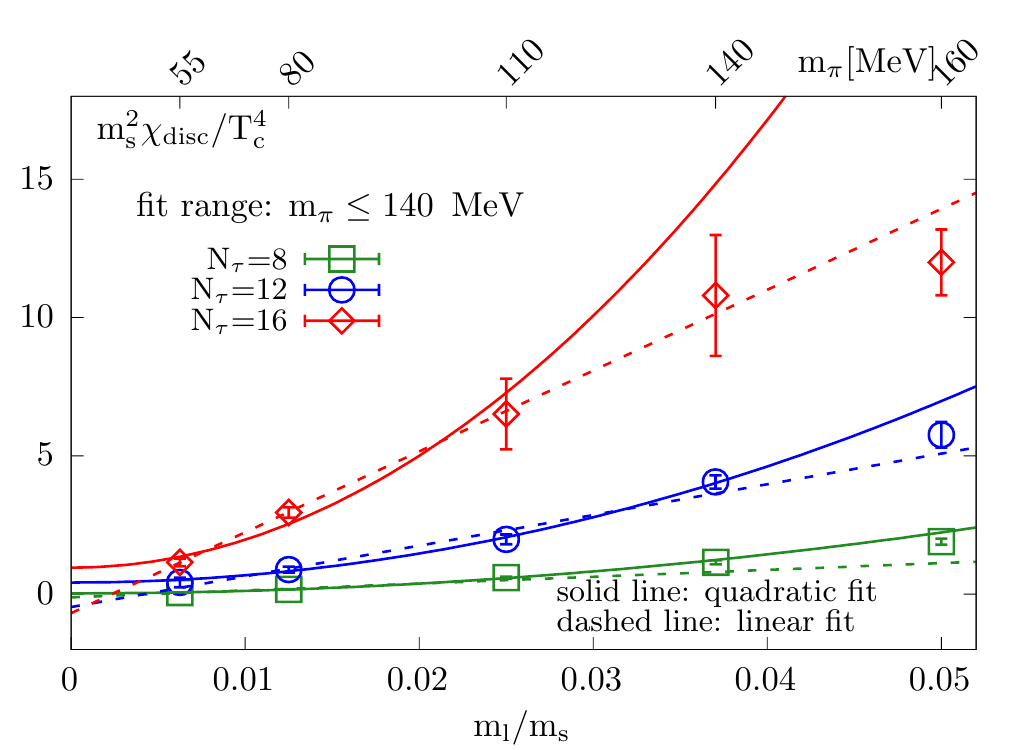}
	\caption{Left: $m_s\pbp/T_c^4$ as a function of quark mass for three lattice spacings with two different fit ansatz. The solid lines denote linear fits in quark mass while the dashed lines denote quadratic fits in quark mass. Right: Same as the left one but for $m_s^2\cdsc/T_c^4$. Here the solid lines denote quadratic fits while the dashed lines represent linear fits in quark mass.}
	\label{fig:sup_su2}
\end{figure*}

In the left panel of~\autoref{fig:sup_su2} we show the quark mass dependence of chiral condensate in detail. We performed linear fits (solid lines) and a quadratic fits (dotted lines) in quark mass to the chiral condensate. It can be clearly seen that the linear fits give a good description of the data and the fit result of chiral condensates at each lattice spacing vanish in the chiral limit. This is in accord with the expectation $Z[\U]$ is an even function of $\ml$ for $T\ge\tc$ due to the restoration of the $Z(2)$ subgroup of $\su$~\footnote{In the staggered discretization formalism the remnant chiral symmetry at nonzero lattice spacing is O(2).}. This leads to the expectation that the $\cdsc$ should be quadratic in quark mass as $\ml \to 0$. As can be seen from the right panel of~\autoref{fig:sup_su2} which shows the $m_s^2\cdsc/T_c^4$ as a function of quark mass for $\nt=8, 12, 16$, the data indeed favors the quadratic dependence of $m_s^2\cdsc/T_c^4$ in quark mass as $\ml \to 0$.
\begin{figure}[!htp]
  \centering
	\includegraphics[width=0.495\textwidth, height=0.25\textheight]{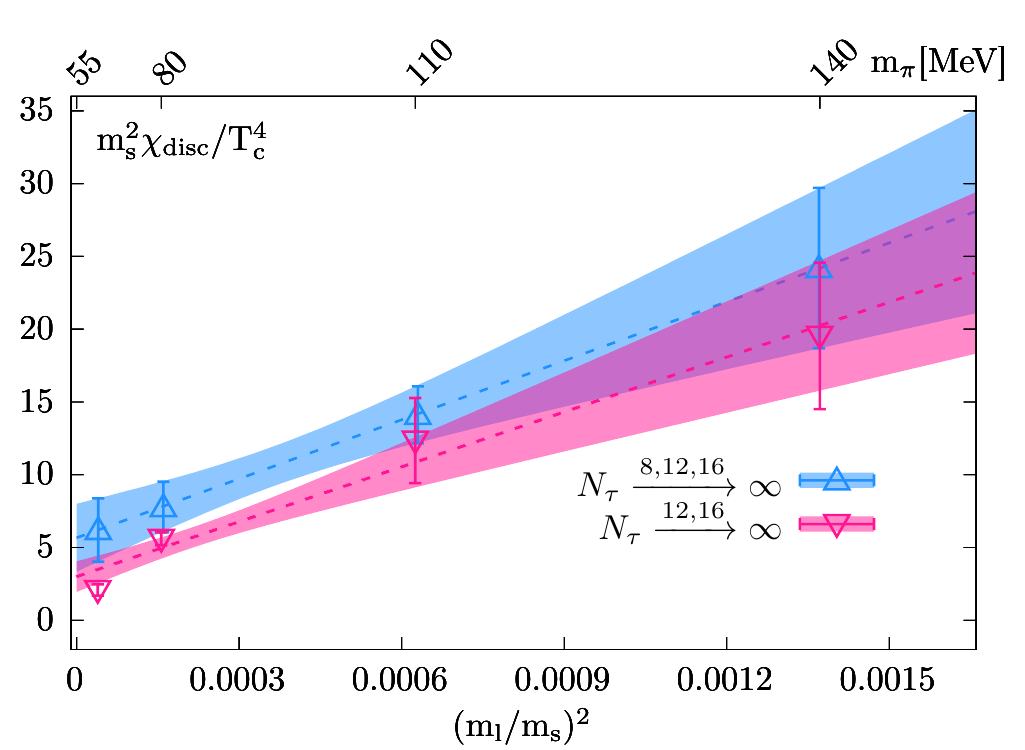}
	\includegraphics[width=0.495\textwidth, height=0.25\textheight]{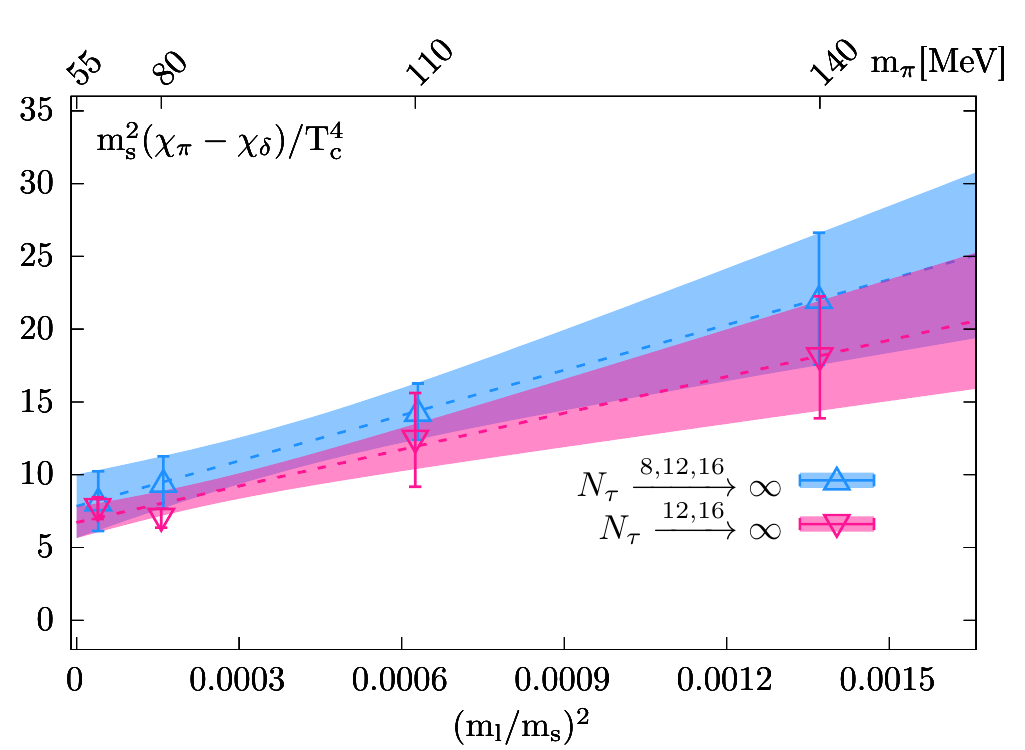}
	\caption{Continuum and chiral extrapolated results for $\cdsc$ (left) and
	$\cpi-\cdl$ (right) at $T\approx205$~MeV. See text for details.}
  \label{fig:cont_chiral_chi}
\end{figure}

In \autoref{fig:cont_chiral_chi} we show the continuum and chiral extrapolated
results for $\cdsc$ and $\cpi-\cdl$. With the additional 2 data points at $m_s/m_l=160$ (or $m_\pi=55$ MeV) on $N_\tau$= 12 and 16,  we follow the same analysis methods as in our previous studies~\cite{Ding:2020xlj}. I.e. using data for $\nt=8, 12, 16$ and
$m_\pi\le140$~MeV, we performed a joint $a, \ml \to 0$ extrapolation of the form
$\cdsc(a,\ml) = \cdsc(0,0) +a_1/\nt^2 + a_2/\nt^4 + \qty(\ml/\ms)^2\qty[ b_0+
b_1/\nt^2 + b_2/\nt^4 ]$. Fits were performed on each bootstrap sample of the data
set. The bootstrap samples were created by randomly choosing data from Gaussian
distributions with means equal to the average values and variances equal to the errors of $\cdsc$. We chose the median value as the final result (depicted by the upward
triangles) and the 68\% percentiles confidence interval of the resulting bootstrap distribution
as the errors (the band labeled by
$N_\tau\xrightarrow{8,12,16}\infty$). Since we used the so-called rooted-staggered
formulation~\cite{Bernard:2006vv}, we also checked that the same $\cdsc(0,0)$ is obtained
within errors by first carrying out the $a\to0$ extrapolations for each $\ml$ and
then performing the $\ml\to0$ extrapolation.
For this purpose, we used the $\nt=12, 16$ data for each of $\ml = \ms/27, \ms/40,
\ms/80, \ms/160$ to obtain $\cdsc(0,\ml)$ by fitting to the ansatz $\cdsc (a,m_{l})=\cdsc
(0,m_{l})+d_{1}/N_{\tau}^{2}$. Then the chiral extrapolation was carried out using
$\cdsc(0,\ml) = \cdsc(0,0) + d_2(\ml/\ms)^2$ based on the continuum estimates of
$\cdsc(0,\ml)$. These extrapolations were done by using the same bootstrap procedure
described before and the final results are indicated with the label
$N_\tau\xrightarrow{12,16}\infty$. The same procedures were followed also
for $\cpi-\cdl$ to obtain its continuum and chiral extrapolated values. After
carrying out continuum and chiral extrapolations we obtained that $\cdsc(0,0)$ is $3.0\pm 1.1$ for the sequential fit and $5.7\pm2.3$ for the joint fit, which is 2-3 $\sigma$ away from 0, while $[\cpi-\cdl](0,0)$ is $6.7\pm 1.1$ for the sequential fit and $7.8\pm2.2$ for the joint fit, which is 4-6 $\sigma$ away from 0. We
find that \autoref{eq:rel} is
satisfied within errors, and $\cdsc$ and $\cpi-\cdl$ are nonvanishing at a confidence level above 95\%. These results are consistent with those obtained without the two additional data points~\cite{Ding:2020xlj}.

\section{Conclusions}
\label{5}
 In this work we establish relations between $\partial^n \rho/\partial m_l^n$ and $C_{n+1}$. Based on these
relations, we present direct computations of
$\partial^n \rho/\partial m_l^n$ employing state-of-the-art lattice QCD techniques. Based on these results we conclude that, in chiral symmetric (2+1)-flavor QCD at
$T\approx1.6\tc$, (i) $\rho(\lda\to0,\ml)$ develops a peaked structure due to
repulsive non-Poisson correlations within small $\lda$; the peak becomes sharper as
$a\to0$, and its amplitude is $\propto\ml^2$. (ii) The underlying presence of this
$\rho(\lda\to0,\ml)$ leads to manifestations of $\ua$ anomaly in $\cpi-\cdl$ and
$\cdsc$. (iii) Axial anomaly remains manifested in $\cpi-\cdl$ and $\cdsc$ even in
the chiral limit. These suggest that for $T \sim1.6\tc$ the microscopic origin of
axial anomaly is driven by the weakly interacting (quasi)instanton gas motivated
$\rho(\lda\to0,\ml\to0)\sim\ml^2\delta(\lda)$, and the chiral phase transition in
(2+1)-flavor QCD is of the three-dimensional $O(4)$ universality class.

The above conclusions are based on the continuum extrapolated lattice QCD
calculations using the (2+1) flavors of staggered fermions. Confirmations of these continuum extrapolated results using other fermion actions, especially using chiral fermions, are needed in future. Even in those future calculations it will be very difficult to directly identify a structure like $\ml^2\delta(\lda)$ in $\rho$ itself
as $\ml\to0$. The formalism developed and techniques presented in this work for directly accessing $\partial^n \rho/\partial m_l^n$ will be essential for those future studies too.

\section*{Acknowledgement}
This material is based upon work supported by the National Natural Science
Foundation of China under Grants No. 11775096, No. 11535012, and No. 11947237; the U.S.
Department of Energy, Office of Science, Office of Nuclear Physics through the Award No. DE-SC0012704; the U.S. Department of Energy,
Office of Science, Office of Nuclear Physics and Office of Advanced Scientific
Computing Research within the framework of Scientific Discovery through Advance
Computing (SciDAC) award Computing the Properties of Matter with Leadership Computing
Resources; and RIKEN Special Postdoctoral Researcher program and JSPS KAKENHI Grant No. JP20K14479.
Computations for this work were carried out on the GPU clusters of the Nuclear
Science Computing Center at Central China Normal University (NSC$^3$), Wuhan, China,
and facilities of the USQCD Collaboration, which are funded by the Office of Science
of the U.S. Department of Energy.
For generating the gauge configurations, the HotQCD software suite was used, and the
eigenvalue measurement code was developed also based on the same software suite. We
are indebted to the HotQCD Collaboration for sharing their software suite with us.

\bibliographystyle{JHEP.bst}
\bibliography{ref.bib}

\end{document}